\documentclass[12pt]{article}
\usepackage{mathrsfs}
\usepackage{amsmath}
\usepackage{mathrsfs}
\usepackage{amssymb}
\usepackage{color}
\usepackage{psfrag}
\usepackage{graphicx}
\usepackage{subfigure}
\usepackage{rotating}
\usepackage{cite}
\usepackage[colorlinks=true,linktocpage=true,linkcolor=blue,citecolor=blue]{hyperref}
\usepackage[section]{placeins}
\usepackage{amsfonts}
\usepackage{ifpdf}
\usepackage[page, toc,titletoc]{appendix}
\ifpdf
\else

\newcommand{\bea}{\begin{eqnarray}}
\newcommand{\eea}{\end{eqnarray}}
\newcommand{\be}{\begin{equation}}
\newcommand{\ee}{\end{equation}}
\renewcommand{\a}{\alpha}
\renewcommand{\b}{\beta}
\newcommand{\g}{\gamma}

\newcommand{\de}{\delta}
\newcommand{\e}{\epsilon}
\newcommand{\la}{\lambda}

\newcommand{\ka}{\kappa}

\newcommand{\half}{\frac{1}{2}}
\newcommand{\pa}{\partial}
\newcommand{\si}{\sigma}
\newcommand{\Si}{\Sigma}
\renewcommand{\t}{\theta}
\newcommand{\nn}{\nonumber}

\newcommand{\mc}{\mathcal}
\newcommand{\f}{\frac}
\newcommand{\ra}{\rightarrow}

\newcommand{\vs}[1]{\vspace{#1 mm}}

\begin{document}
\topmargin 0pt
\oddsidemargin 0mm

\vspace{2mm}

\begin{center}

{\Large \bf {Drag force in strongly coupled, anisotropic plasma at finite chemical potential}}

\vs{10}

{Somdeb Chakraborty\footnote{E-mail: somdeb.chakraborty@saha.ac.in}and Najmul Haque\footnote{E-mail: najmul.haque@saha.ac.in}}

 \vspace{4mm}

{\em
 Theory Division,
 Saha Institute of Nuclear Physics,\\
 1/AF Bidhannagar, Kolkata-700 064, India\\}

\end{center}

\vs{10}

\begin{abstract}
\noindent
We employ methods of gauge/string duality to analyze the drag force on a heavy quark moving through a strongly coupled, anisotropic $\mc{N}=4, SU(N)$ super Yang-Mills plasma in the presence of a finite $U(1)$ chemical potential. We present numerical results valid for any value of the anisotropy parameter and the $U(1)$ charge density and arbitrary direction of the quark velocity with respect to the direction of anisotropy. In the small anisotropy limit we are also able to furnish analytical results.
\end{abstract}

\newpage
\tableofcontents

\section{Introduction} \label{intro}
Recent heavy ion experiments at the Relativistic Heavy Ion Collider (RHIC) \cite{Adams:2005dq,Adcox:2004mh } and the Large Hadron Collider (LHC) have provided precious insights into the rich physics underlying quark-gluon plasma (QGP), simultaneously presenting an enormous challenge to understand the wealth of experimental data from a theoretical standpoint. One of the most fascinating results to emerge out of the data obtained at RHIC is that the quark-gluon plasma formed as an outcome of the heavy ion collisions is dominated by strong coupling effects \cite{Shuryak:2003xe,Shuryak:2004cy}. The strong coupling scenario urges for novel techniques to study such systems, since the conventional field-theoretic approach relies upon perturbative tools and, as such, may not be reliable in the strong coupling domain.  The AdS/CFT correspondence or more generally, the gauge/string duality \cite{Maldacena:1997re, Witten:1998qj,Gubser:1998bc,Aharony:1999ti },  on the other hand, seems to be tailor-made to handle such 
situations. The best understood and most investigated example of this duality conjectures an equivalence between type IIB string theory on $AdS_5 \times S^5$ and $\mc{N}=4, SU(N)$ super Yang-Mills (SYM) theory living on the $(3+1)$-dimensional boundary of the $AdS_5$ space. It  enables us to extract information about the strongly coupled SYM theory by studying the weakly coupled string dual. Since then the duality has been generalized in various directions to embrace a larger variety of gauge theories under its scope.\\
One of the frontier areas where the duality has been applied with reasonable success is the physics of quark-gluon plasma. Collisions of heavy nuclei are usually marked by a non-zero impact parameter which results in a pressure gradient building up in the region where the two colliding nuclei overlap. The initial coordinate-space anisotropy is transformed into an observed momentum-space anisotropy, {\it via} the interactions between the produced particles finally leading to an anisotropic particle distribution. The relativistic ideal hydrodynamical models tasted early success in explaining soft collective flow and hadronic spectra at RHIC. This provided empirical evidence in favor of fast thermalization and isotropization - at time scales $\sim 0.5$ fm. In an attempt to make better agreement with experimental results this was subsequently generalized to incorporate viscous effects, which, however,  predicted the presence of a sizable pressure anisotropy. Hydrodynamical simulations suggest that the transverse 
pressure exceeds the longitudinal (along the beam direction) one with the disparity being the most conspicuous for time $ \leq 2$ fm. Hence, to describe a realistic situation, it is desirable to take into account the presence of anisotropy in the system. Though in a real plasma the strength of anisotropy is expected to die away with time, in many cases it suffices to consider the anisotropy to be constant over the time-scale of interest. Encouraged by the field-theoretical studies on anisotropic QGP \cite{Ryblewski:2011aq, Martinez:2010sd, Ryblewski:2010bs, Martinez:2010sc}, there have also been a  spate of activities related to studying strongly coupled, anisotropic QGP {\it via} the gauge/string duality \cite{Maldacena:1999mh, Alishahiha:1999ci, Harmark:1999rb,Cai:1999aw, Cai:2000hn, Chakraborty:2011gn, Chakraborty:2012pc }. Recently, Mateos and Trancanelli \cite{Mateos:2011ix, Mateos:2011tv} constructed a gravity solution firmly embedded in type IIB superstring theory that is dual to a topologically 
deformed SYM theory, where the topological deformation injects anisotropy into the theory. Although the source of anisotropy in such a model is radically different from that in actual QGP, nevertheless, one expects this model to encapsulate the telltale signatures of anisotropy on various QGP observables. Various  quantities related to heavy quark observables were subsequently computed in this model \cite{Rebhan:2011vd, Chernicoff:2012iq, Giataganas:2012zy, Chernicoff:2012gu, Rebhan:2012bw, Fadafan:2012qu, Chernicoff:2012bu, Muller:2012uu, Patino:2012py, Chakraborty:2012dt, Wu:2013qja, Finazzo:2013rqy, Fadafan:2013bva, Chakrabortty:2013kra, Jahnke:2013rca, Giataganas:2013zaa, Ali-Akbari:2014xea, Ali-Akbari:2014nua}. An interesting scenario to consider is when the QGP formed is endowed with a finite baryon number density. At RHIC, the QGP is assumed to carry a finite, albeit small, baryon number density and is characterized by a high temperature. Forthcoming collider programs at the Facilty for Antiproton and 
Ion Research (Gesellschaft f\"{u}r Schwerionenforschung, Germany) and the Nuclotron-based Ion Collider fAcility (Joint Institute for Nuclear Research, Russia) are expected to thoroughly explore plasma characterized by a high baryon number density and moderate temperature. To study such systems using the AdS/CFT toolkit, one needs to introduce a $U(1)$ chemical potential in the dual gravity description.  The easiest way to implement this is to consider black branes carrying $U(1)$ $R$-charge. Recently, the authors of \cite{Cheng:2014sxa, Cheng:2014qia} presented a type IIB solution dual to $\mc{N}=4$, anisotropic SYM plasma and carrying a non-zero $U(1)$ charge. We use this supergravity background to compute the drag force experienced by a heavy quark moving in an anisotropic, charged thermal medium.  While the drag force in both anisotropic \cite{Chernicoff:2012iq} and charged \cite{Caceres:2006dj, Fadafan:2008uv} plasma have been computed previously, it is of interest to see the outcome of the interplay 
between 
the two parameters. Motivated by this, we investigate in this paper  the drag force experienced by an energetic quark moving through hot, anisotropic, charged plasma and uncover some rather remarkable results.\\
The paper is organized as follows. In section \ref{dual} we elaborate upon  the gauge theory that we wish to consider and the dual gravity description which was first studied in \cite{Cheng:2014sxa, Cheng:2014qia}. In section \ref{drag} we study the drag force acting on a massive quark as it moves through the plasma and present numerical results.  Further, by restricting to the small anisotropy and small charge density regime we are able to furnish analytical results for the drag force. Finally, we conclude with a discussion of our results in section \ref{conc}.

\section{The dual geometry} \label{dual}
\begin{figure}[tbh]
\begin{center}
\subfigure[]{
\includegraphics[width=7.2cm,height=5.6cm, angle=-0]{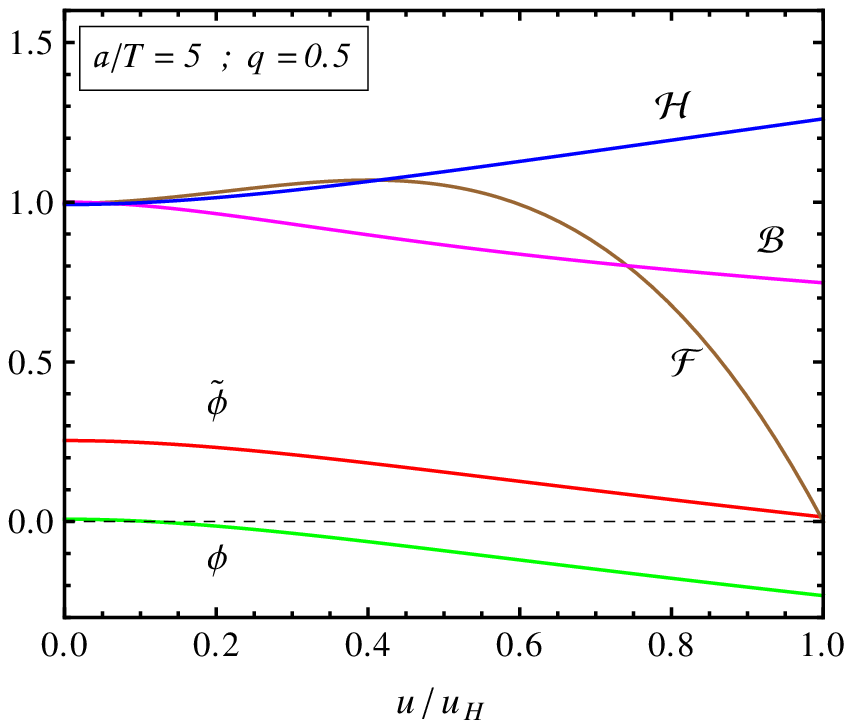}}
\hspace{5mm}
\subfigure[]{
\includegraphics[width=7.2cm,height=5.7cm, angle=-0]{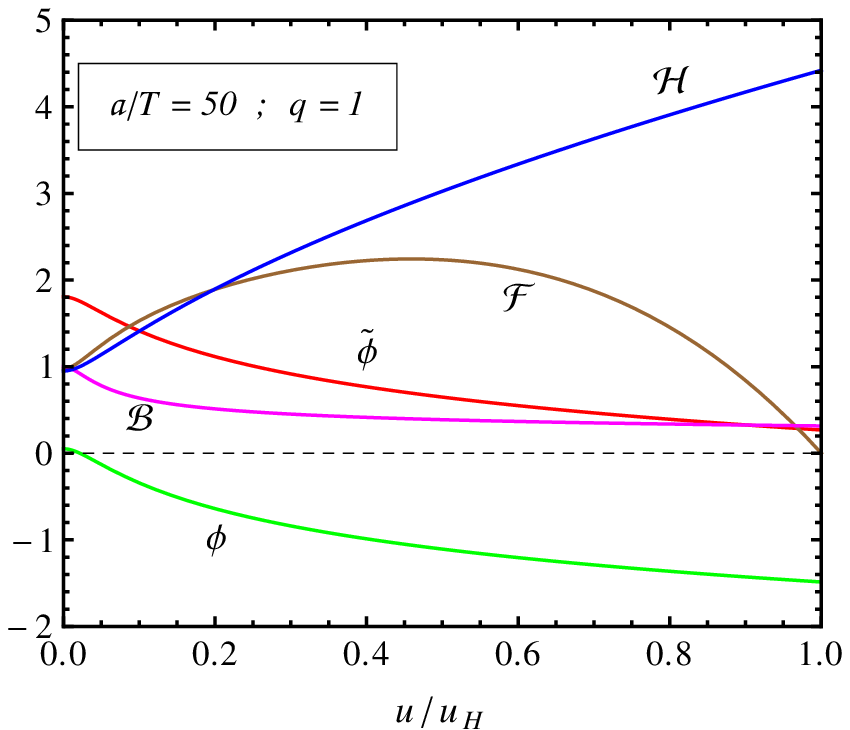}}
\caption{\label{metricfn} \small{(a) shows numerical results for $\mc{F}, \mc{B}, \mc{H}, \phi$ and $\tilde{\phi}$ for $a/T=5, q=0.5$. In (b) the same quantities have been plotted with $a/T=50, q=1.0$.}}
\end{center}
\end{figure}

In \cite{Mateos:2011ix, Mateos:2011tv} Mateos and Trancanelli obtained a type IIB supergravity solution dual to a topologically deformed, thermal  $\mc{N}=4$, $SU(N)$  SYM theory at large 't Hooft coupling $\la=g_{\text{YM}}^2 N$. The gauge theory of interest is a deformed cousin of the $\mc{N}=4$ SYM theory where the deformation is effected by a topological term,
\be 
S=S_{\text{SYM}}+\f{1}{8 \pi^2}\int \t (x^3) \text{Tr} F \wedge F
\ee
where $\{t,x^1,x^2,x^3\}$ define the gauge theory coordinates and the $\t$-term, that depends upon only one of the gauge theory space coordinates (in this case $x^3$), is responsible for injecting anisotropy into the theory thereby spoiling the $SO(3)$ symmetry. Now a remnant $SO(2)$ symmetry is preserved only in the $x^1$-$x^2$ plane. While trying to relate the situation to heavy ion collisions it is then only natural to identify the $x^3$ direction with the beam direction. Aiming for a more realistic model, this was recently generalized to encompass the case where the plasma is characterized by a finite $U(1)$ charge density  \cite{Cheng:2014sxa, Cheng:2014qia}. In the dual picture one considers a Einstein-Maxwell-dilaton-axion system in $10$-dimensional type IIB supergravity compactified upon $S^5$.  The non-linear Kaluza-Klein reduction of $10$-dimensional supergravity on $5$-dimensions results in an Abelian gauge field. The relevant $5$-dimensional action is given in the Einstein frame by,
\be 
S=\f{1}{2\ka^2}\int_{\mc{M}}\sqrt{-g}\left( \mc{R}+12-\half (\pa \phi)^2-\half e^{2 \phi}(\pa \chi)^2 -\f{1}{4}F_{MN}F^{MN}\right)+\f{1}{2\ka^2}\int _{\pa \mc{M}}\sqrt{-\g}2K
\ee
where $2\ka^2 \equiv 16 \pi G_5=\f{8 \pi^2}{N^2}$\footnote{We have set the radius $R$ of $S^5$, which is a constant in the Einstein frame, to be unity.} is the gravitational coupling in $5$-dimensions, $F_{MN}=\pa_{[M} A_{N]}$. $\phi$ and $\chi$ are the dilatonic and the axionic excitations  respectively and the last term is the usual Gibbons-Hawking surface term defined on the $4$-dimensional boundary $\pa \mc{M}$ of the $5$-dimensional manifold $\mc{M}$ and $\g_{\mu \nu}$ being the metric on $\pa \mc{M}$. The $5$-dimensional solution, given in the string frame, is taken to have the form,
\be \label{Sin}
ds^{2}=\f{1}{u^2}\left(-\mc{FB}dt^{2}+(dx^{1})^{2}+(dx^{2})^{2}+\mc{H}(dx^{3})^{2}+\f{du^{2}}{\mc{F}}\right)\equiv G_{MN}dX^M dX^N,
\ee
\be 
\chi=ax^3, \qquad \phi=\phi(u), \qquad \mc{H}=e^{-\phi}
\ee
where $a$ is a measure of anisotropy. The axion $\chi$ is dual to the $\t$-term in the gauge theory. $u$ indicates the radial coordinate with the gauge theory living at $u=0$. $X^M$, $\{M=0,...,9\}$ are the $10$-dimensional coordinates. The gauge theory coordinates are defined by $t=X^0$ and $x^i=X^i (u=0)$ $\{i=1,2,3\}$. The metric components $\mc{F},\mc{B},\mc{H}$ are only functions of the radial coordinate $u$. Evidently, in an isotropic  theory one sets $\mc{H}=1$. The black brane horizon is defined by the radial coordinate $u_H$ such that $\mc{F}(u_H)=0$. We further demand that asymptotically the metric becomes $AdS_5$ which imposes the constraints: $\phi(0)=0, \mc{F}(0)=\mc{B}(0)=\mc{H}(0)=1$. Interestingly, the authors in \cite{Cheng:2014qia} found two classes of solution, which they termed as prolate and oblate depending upon whether $\mc{H}(u_H) >1$ or $\mc{H}(u_H) <1$. This, in turn, corresponds to whether $a$ is real or imaginary. In this paper, we shall only work with the prolate solution 
implying $a \in \mathbb{R}$. The temperature $T$ and the entropy density $s$ can easily be found as,
\bea 
T&=&-\frac{1}{4 \pi} \mc{F}'_{H}\sqrt{\mc{B}_H}\\ \nn
&=& \sqrt{\mc{B}_H}\left(\frac{e^{-\frac{\phi_H}{2}}}{16 \pi u_H} (16+a^2 e^{7 \frac{\phi_H}{2}}u_H^2)-\frac{e^{2 \phi_H}Q^2 u_H^5}{24 \pi}\right)\\
s&=&\frac{N^2 e^{-\frac{5}{4}\phi_H}}{2 \pi u_H^3}
\eea
where $Q$ is a constant related to the $U(1)$ charge density on the gauge theory side and the prime indicates derivative with respect to the radial coordinate. In the above we have introduced the notations: $\mc{F}(u_H) \equiv \mc{F}_{H}$ and so on. The gauge field has only one non-vanishing component,
\be 
A_t(u)=-\int_{u_H}^{u}Q \sqrt{\mc{B}}e^{\frac{3}{4}\phi}udu
\ee
supplemented by the constraint that it vanishes on the horizon, i.e., $A_t(u_H)=0$
whence the  $U(1)$ chemical potential is obtained as
\be 
\mu=\int_{0}^{u_H}Q \sqrt{\mc{B}}e^{\frac{3}{4}\phi}udu.
\ee
Note that setting $a=0$ we recover the temperature of the RN-AdS black brane temperature,
\be 
T_{a=0}=\frac{1}{2 \pi u_H} (2-q^2)
\ee
where we have defined $q \equiv \f{u_H^3 Q}{2\sqrt{3}}$. As $q \rightarrow \sqrt{2}$ we approach the extremal RN-AdS solution which sets an upper bound on $q$ as $q^2_{\text{max}}=2$. It has been argued in \cite{Cheng:2014qia} that as one switches on a non-trivial dilaton profile with $a>0$ the horizon of the anisotropic RN-AdS black brane is always greater than its isotropic cousin signifying that the extremal limit can not be accessed. The metric functions $\mc{F}, \mc{B}, \mc{H}$ and the dilaton profile $\phi(u)$ can be found out analytically only in the small anisotropy limit by perturbing around the $a=0$ solution. Otherwise, they can be obtained numerically for any value of $a$. In Fig.\ref{metricfn} we provide numerical plots of these functions for some sample values of the parameters\footnote{See Appendix \ref{app} for details of the numerical scheme followed.}.
For future reference, let us also discuss the case when both $a$ and $q$ are kept small, whence, it is possible to furnish some analytical results. In this regime of the parameter space let us write the relevant functions as,
\bea
\mc{F}(u)&=&\mc{F}_0(u)+a^2 \left( \hat{\mc{F}}_0(u) +\hat{\mc{F}}_2(u)q^2 +\mc{O}(q)^4\right)+\mc{O}(a)^4, \nn \\
\mc{B}(u)&=&\mc{B}_0(u)+a^2 \left( \hat{\mc{B}}_0(u) +\hat{\mc{B}}_2(u)q^2 +\mc{O}(q)^4\right)+\mc{O}(a)^4, \nn \\
\phi(u)&=&\phi_0(u)+a^2 \left( \hat{\phi}_0(u) +\hat{\phi}_2(u)q^2+ \mc{O}(q)^4\right) +\mc{O}(a)^4
\eea
where
\bea
\mc{F}_0(u)&=&1-\left( \f{u}{u_H}\right)^4+q^2\left[\left(\f{u}{u_H}\right)^6-\left(\f{u}{u_H}\right)^4 \right], \nn \\
\hat{\mc{F}}_0(u)&=&\f{1}{24u_H^2}\left[8u^2(u_H^2-u^2)-10u^4\log2+(3u_H^4+7u^4)\log\left( 1+\f{u^2}{u_H^2}\right) \right], \nn \\
\hat{\mc{F}}_2(u)&=& \f{1}{24u_H^4(u^2+u_H^2)}\left[ 7u^8+6u^2u_H^6+u^4u_H^4(25\log2-12)+u^6u_H^2(25\log2-1)\right. \nn \\
&\quad&\left. -(u^2+u_H^2)(12u^6+7u^4u_H^2+6u_H^6) \log\left(1+\f{u^2}{u_H^2}\right) \right], \nn \\
\mc{B}_0(u)&=&1, \nn \\
\hat{\mc{B}}_0(u)&=&-\f{u_H^2}{24}\left[ \f{10u^2}{u^2+u_H^2}+\log\left(1+\f{u^2}{u_H^2}\right) \right], \nn \\
\hat{\mc{B}}_2(u)&=&\f{1}{24}\left[-\f{u^2(11u^4+3u^2u_H^2+2u_H^4)}{(u^2+u_H^2)^2} +2u_H^2\log\left(1+\f{u^2}{u_H^2}\right) \right], \nn \\
\phi_0(u)&=&0, \nn \\
\hat{\phi}_0(u)&=&-\f{u_H^2}{4}\log\left(1+\f{u^2}{u_H^2}\right) , \nn \\
\hat{\phi}_2(u)&=&\f{1}{4}\left[-2u^2+\f{u^4}{u^2+u_H^2}+2u_H^2 \log\left(1+\f{u^2}{u_H^2}\right)  \right].
\eea
The gauge field takes the form,
\bea
A_t=\f{q}{8 \sqrt{3}u_H^3}\left[24(u_H^2-u^2)+5a^2u_H^2\left(u^2 \log\left(1+\f{u^2}{u_H^2}\right)  -u_H^2 \log2\right) \right]
\eea
with the corresponding chemical potential
\be 
\mu=\f{q (24-5a^2u_H^2\log2)}{8\sqrt{3}u_H}.
\ee
The temperature now reads,
\be 
T=\f{2-q^2}{2\pi u_H}+a^2\f{u_H}{96 \pi}\left[10\log2-4+5(3+\log2)q^2 \right] +\mc{O}(a^4).
\ee
Note that for a fixed $u_H$, for $a=0$, $T$ decreases with increasing $q$, whereas by setting $q=0$ one finds that $T$ increases with rise in $a$. 
\begin{figure}[tbh]
\begin{center}
\subfigure[]{
\includegraphics[width=7.2cm,height=5.6cm, angle=-0]{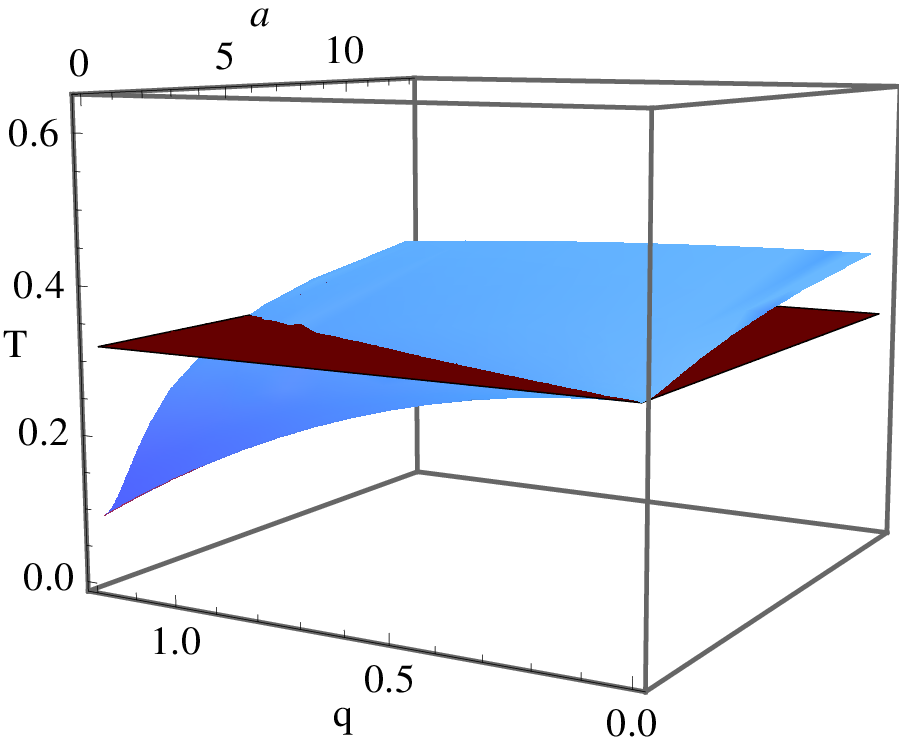}}
\hspace{5mm}
\subfigure[]{
\includegraphics[width=6.2cm,height=5.cm, angle=-0]{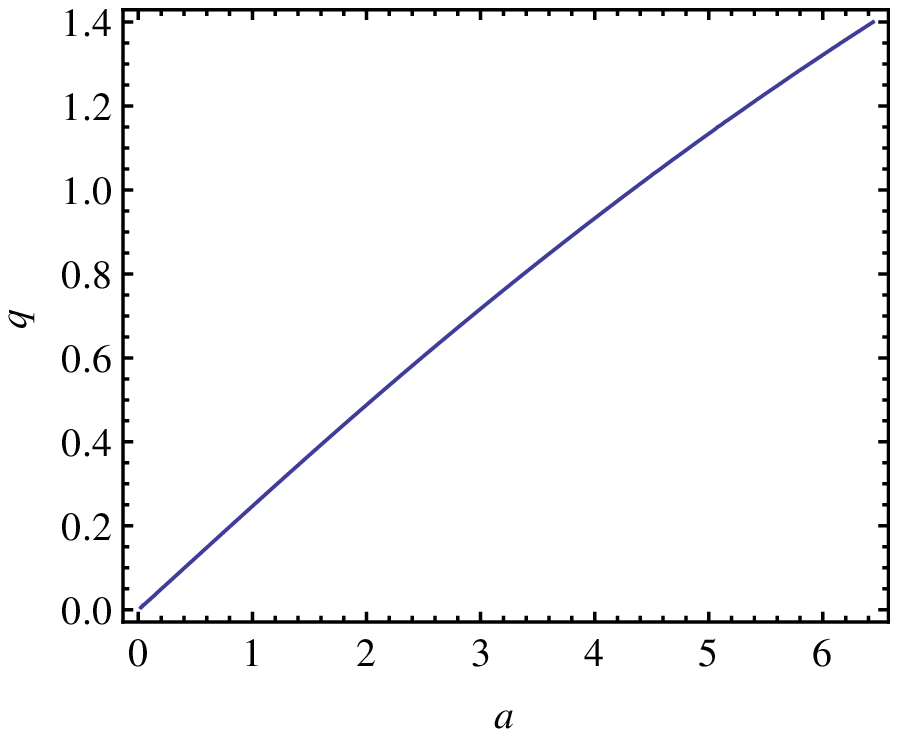}}
\caption{\label{temp} \small{ (a) shows the temperature of the AdS-BH (red surface) and that of the anisotropic, charged black brane (blue surface) with varying $q$ and $a$. (b) shows the line in the $a$-$q$ plane along which the temperature of the anisotropic, charged black brane equals the AdS-BH temperature.}}
\end{center}
\end{figure}
An interesting feature of the charged, anisotropic black brane solution is that when both these parameters are switched on, there exists a regime in the parameter space spanned by $a$ and $q$ where the brane temperature is less than that of an isotropic, uncharged solution, i.e., AdS-black hole (AdS-BH). In the opposite regime, the temperature is greater than that of a AdS-BH. The two regimes are separated by a line in the $a$-$q$ plane along which the transition takes place and the temperature equals that of the AdS-BH  geometry.  This has been shown clearly in Fig.\ref{temp}. For small $a$ and $q$ the functional form of the transition line is given by,
\be
q(a)=\f{1}{4}\sqrt{\f{10\log2-4}{3}}a+\f{5}{384}(3+\log2)\sqrt{\f{10\log2-4}{3}}a^3+\mc{O}(a)^5.
\ee

\section{Drag force} \label{drag}
In this section we evaluate the drag force felt by a heavy quark moving through a strongly coupled, anisotropic, charged plasma. The analysis of drag force on a moving quark was first done in the seminal works \cite{Herzog:2006gh, Gubser:2006bz}. Since then the collisional energy loss has been computed in various other contexts using the gauge/string duality. Following the holographic prescription suggested in \cite{Herzog:2006gh, Gubser:2006bz} we consider a heavy quark of mass $m$ moving through the thermal bath with a velocity $\vec{v}$. A simple phenomenological model that encapsulates the dynamics of the quark is given by,
\be 
\f{d\vec{p}}{dt}=-\mu \vec{p} + \vec{F}
\ee
where $\mu$ denotes the coefficient of drag and $\vec{F}$ is an externally applied force. $\vec{p} \equiv m \vec{v}$ is the quark momentum. The steady state condition is then defined by $\mu\vec{p}=\vec{F}$ when the quark moves with constant velocity. It is worth noting at this stage, that by virtue of the anisotropy of the medium, the drag coefficient $\mu$ is no longer a number but is now matrix-valued which we write as $\mu=\{\mu_1,\mu_2,\mu_3 \}$. Moreover, the $SO(2)$ invariance in the $x^1$-$x^2$ plane dictates that $\mu_1=\mu_2$. In the dual gravity picture a fundamental quark corresponds to a string embedded in the geometry (Eq.\ref{Sin}), hanging from the boundary at $u=0$ towards the horizon at $u=u_H$. The string dynamics is captured by the Nambu-Goto string world-sheet action,
\be \label{NGS}
S_{\text{NG}}=-\f{1}{2 \pi \a'}\int d\vec{\xi}\sqrt{-\text{det}g_{\a \b}}\equiv \int d\vec{\xi} \mc{L}
\ee
where $g_{\a \b}$ denotes the pull-back,
\be 
g_{\a \b}=G_{MN}\pa_{\a}X^M\pa_{\b}X^N
\ee
and $\vec{\xi}\equiv \{\xi^0,\xi^1\}\equiv\{\tau,\si\}$ are the coordinates defined on the string world-sheet. The space-time momentum flow is given by,
\be 
\Pi_M=\f{\pa \mc{L}}{\pa(\pa_{\si}X^M)}.
\ee
As alluded to earlier, the quark is held in steady state by some external force. Physically this can be realized by turning on an electric field $F_{MN}=\pa_{[M}A_{N]}$ on the D$7$-brane world-volume, to which the end-point of the fundamental probe string is attached. This implies that the action (Eq.\ref{NGS}) is now modified by the addition of a boundary term
\be 
S=S_{\text{NG}}+S_{\text{b}}
\ee
where
\be 
S_{\text{b}}=-\int_{\pa \Si} d\tau A_M \pa_{\tau}A^M=-\half \int_{\pa \Si}d\tau F_{MN}X^M \pa_{\tau}X^{N}.
\ee
While the boundary term does not affect the dynamics of the string in the bulk, it modifies the boundary conditions that the string end-point has to satisfy. Setting $\de S=0$ results in the boundary condition,
\be \label{bc}
\Pi_M+F_{MN}\pa_{\tau}X^N{\Big{|}_{\pa\Si}}=0.
\ee
To proceed further, we need to fix a convenient gauge. First note that the string does not move along the directions of $S^5$ so that its coordinates on $S^5$ are fixed. Further, rotational invariance in the $x^1$-$x^2$ plane allows us the freedom to set $x^2=0$. Now we employ the gauge-fixing as $\vec{\xi}=\{t,u\}$ and consider the string end-point on the boundary (or, equivalently, the quark) to be moving at an angle $\psi$ to the $x^3$ direction. We focus on string embedding of the form,
\bea
X^1(t,u)&=&(vt+\rho(u))\sin \psi, \nn \\
X^3(t,u)&=&(vt+\zeta(u)) \cos \psi.
\eea
The relevant boundary conditions are,
\bea 
X^1(t,0)&\equiv&x^1(t)=vt \sin \psi, \nn \\
X^3(t,0)&\equiv&x^3(t)=vt \cos \psi
\eea
which is mapped to the trajectory of the quark on the boundary gauge theory. With this choice, the Lagrangian takes the explicit form\footnote{We have suppressed the factor $1/2 \pi \a'$  for the time being. We shall reinstate this factor at the end.},
\be
\mc{L}=-\sqrt{\f{\mc{BF}+\sin^2\psi(\mc{B}\mc{F}^2 \rho'^2-v^2)+\mc{H}\cos^2\psi\left(\mc{B}\mc{F}^2\zeta'^2-v^2-\mc{F}v^2(\rho'^2-\zeta'^2)^2\sin^2\psi \right)}{\mc{F}u^4}}.
\ee
It is now straight forward to compute the rate at which the energy and the momentum is transferred down the string towards the horizon,
\bea \label{pi}
-\Pi_0&=& \f{1}{\mc{L}u^4}\mc{BF}v\left[\rho'\sin^2 \psi+\mc{H}\zeta'\cos^2\psi\right], \nn \\
\Pi_1&=& \f{1}{\mc{L}u^4}\left[\mc{BF}\rho'+\mc{H}v^2(\zeta'-\rho')\cos^2\psi \right]\sin \psi, \nn \\
\Pi_3&=& \f{1}{\mc{L}u^4}\mc{H}\left[\mc{BF}\zeta'+v^2(\rho'-\zeta') \sin^2\psi\right]\cos \psi.
\eea
Correspondingly, the boundary conditions given in Eq.\ref{bc} assume the form,
\be 
\Pi_1=F_1, \qquad \Pi_3=F_3, \qquad -\Pi_0=F_1 v \sin \psi+F_3 v \cos \psi
\ee
where $\vec{F}=\{F_1,F_3\}$ is the force needed to maintain the steady motion of the string. Intuitively, we expect the string to trail behind the quark which is ensured only when $\rho'$ and $\zeta'$ are negative. This, in turn, signifies that the direction of the energy and the momentum flow is from $u=0$ to $u=u_H$. It was argued in \cite{Chernicoff:2012iq} that the string does not trail just below (along the radial direction) the quark - rather there is a misalignment between the transverse position of the quark and the trailing string and the same picture emerges here. To evaluate the string profile we rearrange Eq.\ref{pi} to obtain, 
\bea \label{prime}
\rho'=\pm \sqrt{\f{\mc{H}}{\mc{B}}}\f{v}{\mc{F}}\f{P_1}{\sqrt{P_1 P_3-N}}, \nn \\
\zeta'=\pm \f{v}{\mc{F}\sqrt{\mc{BH}}}\f{P_3}{\sqrt{P_1 P_3-N}}
\eea
where we have introduced the following quantities,
\bea
P_1&=&-\Pi_1\left[\mc{BF}\csc \psi-v^2\sin \psi \right]+\Pi_3 v^2 \cos \psi, \nn \\
P_3&=&-\Pi_3\left[\mc{BF}\sec \psi-\mc{H}v^2\cos \psi \right]+\Pi_1 \mc{H} v^2 \sin \psi, \nn \\
N&=&\f{\mc{BF}\csc \psi \sec \psi}{u^4}\left[\Pi_1\Pi_3 u^4-\mc{H}v^2 \cos \psi \sin \psi \right]\left[\mc{BF}-v^2(\mc{H}\cos^2 \psi +\sin^2 \psi) \right].
\eea
To evaluate the momenta we closely follow the argument developed in \cite{Chernicoff:2012iq}. First note that the quantity $(P_1 P_3-N)_{\big{|}u=0,u_H}>0$. However, generically, $P_1 P_3-N$ can become negative in some range $u_1 <\delta u < u_2$ where $0< u_1 < u_2<u_H$ and $(P_1 P_3-N)_{\big{|}u=u_1,u_2}=0$. Thus, within $\delta u$ both $\rho'$ and $\zeta'$ become imaginary-valued which is unacceptable. To circumvent this pathology, we demand that $P_1 P_3-N \geq0$ in the range $0 \leq u \leq u_H$. This implies that the two zeros of $P_1 P_3-N$ at $u_1,u_2$ coincide. We denote the radial position of this zero by $u_c$. In such a scenario, $P_1 P_3-N$ is always real and positive barring at $u=u_c$ where we demand that the numerators in Eq.\ref{prime} also vanish. Then, both $\rho'$ and $\zeta'$ are smooth and negative in the range $0 \leq u \leq u_H$ provided that in Eq.\ref{prime} the positive sign is chosen for $u < u_c$ and the negative sign is chosen for $u>u_c$. The simultaneous vanishing of the 
numerator and the denominator has the consequence that the zeros of $P_1, P_3$ and the two zeros of $N$ coincide which permits us to solve for the two momenta independently. First of all, setting the second term in $P_1 P_3-N$ to be zero we find $u_c$
\be \label{uc}
\mc{B}_c\mc{F}_c-v^2(\mc{H}_c\cos^2 \psi +\sin^2 \psi)=0
\ee
where $\mc{B}_c \equiv \mc{B}(u_c)$ and so on. Using Eq.\ref{uc} in either $P_1$ or $P_3$ yields a relation between the two momenta
\be
\Pi_1 \mc{H}_c \cos \psi=\Pi_3 \sin \psi.
\ee
Finally, setting the first term in $P_1P_3-Q$ to zero yields the exact expressions for the two momenta
\bea
\Pi_1=\f{v \sin \psi}{u_c^2}, \qquad \Pi_3=\f{\mc{H}_c v \cos \psi}{u_c^2}.
\eea
Hence, the drag force $\vec{F}$ exerted by the medium upon the massive quark moving with constant velocity $\vec{v}=v\{\sin \psi, \cos \psi\}$ is given by\footnote{Strictly speaking, the external force $\vec{F}$ is the negative of the drag force. But we shall be sloppy in our convention and refer to $\vec{F}$ itself to be the drag force.} 
\be 
\vec{F}=\{F_1,F_3\}=\{\Pi_1, \Pi_3\}=\f{\sqrt{\la}}{2\pi}\f{v}{u_c^2}\{\sin \psi, \mc{H}_c \cos \psi \}
\ee
Also note that we have reinstated the factor $1/2 \pi \a'=\sqrt{\la}/2\pi$ in our expression invoking, in the process, the gauge/string dictionary. Further, it is to be observed that the curious feature of the drag force not being aligned with the velocity as found in the anisotropic medium in \cite{Chernicoff:2012iq}, continues to hold here, as expected. In the simplifying limit $a \ra 0$ and $q \ra 0$ the expression above reduces to the drag force in $\mc{N}=4$ SYM plasma,
\be 
\vec{F}_{a=0,q=0}=\{F_1,F_3\}=F_0\{ \sin \psi, \cos \psi\}
\ee
where
\be 
F_0=\f{\pi}{2}\sqrt{\la}T^2\f{v}{\sqrt{1-v^2}}
\ee
and we have used,
\be 
u_c^2=\f{\sqrt{1-v^2}}{\pi^2 T^2}.
\ee

\subsection{Results}
\begin{figure}[tbh] 
\begin{center}
\begin{subfigure}[]{
\includegraphics[width=7.3cm,height=6.1cm, angle=-0]{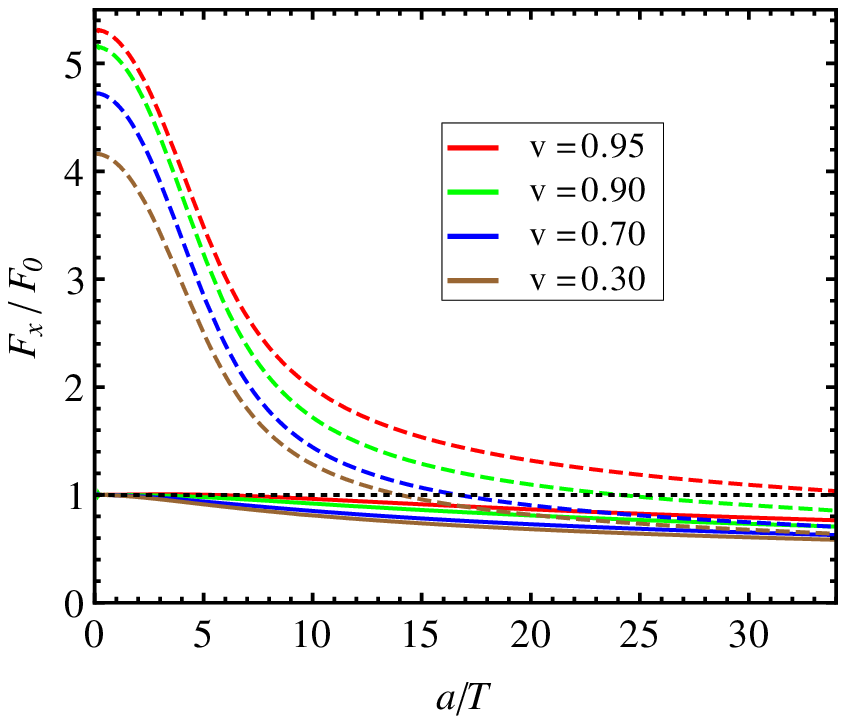}}
\label{Fax}
\end{subfigure}
\hspace{5mm}
\begin{subfigure}[]{
\includegraphics[width=7.5cm,height=6.3cm, angle=-0]{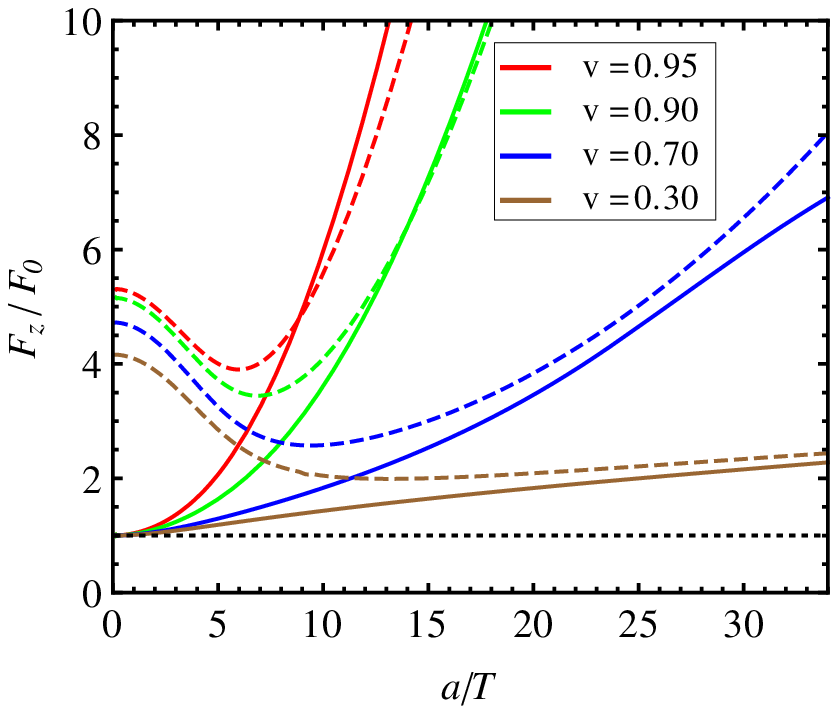}}
\label{Faz}
\end{subfigure}
\caption{\label{Fa} \small{Variation of $F_x$ (a) and $F_z$ (b) with $a/T$ when $q=0,1$ at different values of $v=0.30,0.70,0.90,0.95$.}}
\end{center}
\end{figure}
\begin{figure}[tbh] 
\begin{center}
\subfigure[]{
\includegraphics[width=7.3cm,height=6.1cm, angle=-0]{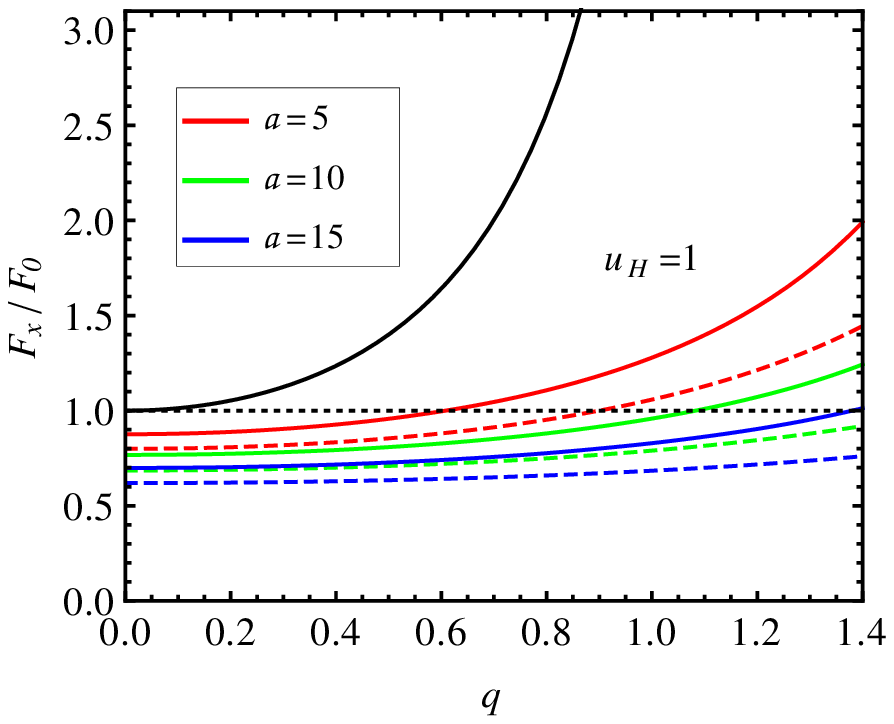}}
\hspace{5mm}
\subfigure[]{
\includegraphics[width=7.4cm,height=6.2cm, angle=-0]{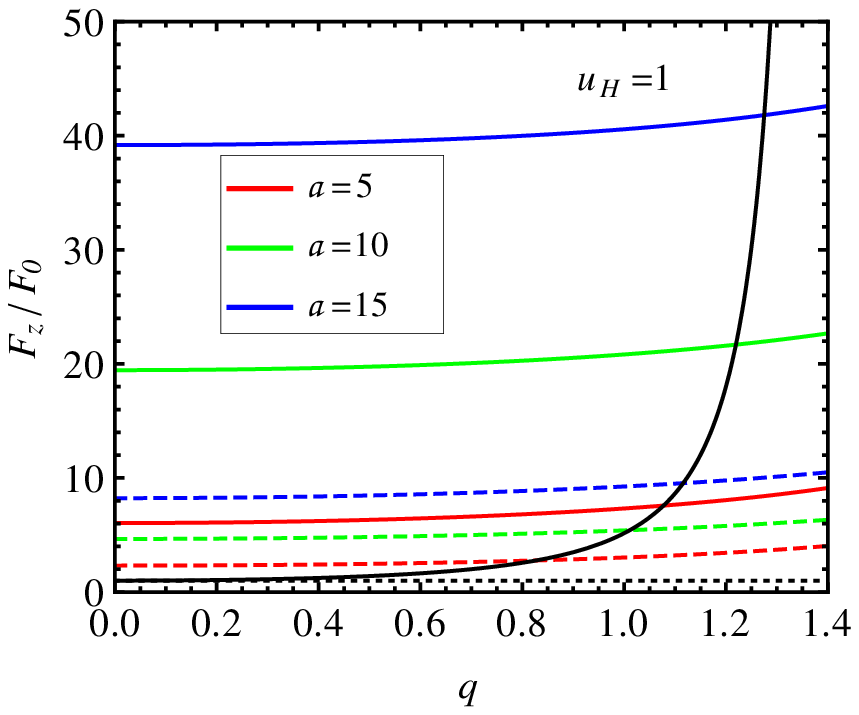}}
\caption{\label{Fq} \small{Variation of $F_x$ (a) and $F_z$ (b) with $q$ when $a$ is fixed at $a=5,10,15$ and $v=0.7,0.9$.}}
\end{center}
\end{figure}
Having developed the requisite formalism, we shall now provide the numerical results. First of all, note that the drag force $\vec{F}$ depends upon the temperature $T$, the velocity $\vec{v}$ of the massive quark, defined by its magnitude $v$ and direction, $\psi$ with respect to the anisotropic axis $x^3$, the strength of the anisotropy parameter $a$ and also the charge density $q$. Thus, $\vec{F}$ has a five parameter dependence, which we denote by $F(T,v,\psi,a,q)$. In particular, we shall use the notation $F_x=F(T,v,\pi/2,a,q)$ and $F_z=F(T,v,0,a,q)$.\\
In Fig.\ref{Fa} we provide the variation of both $F_x$ (Fig.\ref{Fa}(a)) and $F_z$ (Fig.\ref{Fa}(b)) with the anisotropy parameter $a$ (scaled by the temperature $T$) when the charge density $q$ and the velocity $v$ are held fixed. The continuous lines stand for $q=0$ whereas the dashed lines stand for $q=1$. Moreover, we have considered four different values of the velocity $v=0.30,0.70,0.90,0.95$ as is evident from the figure. Further, for convenience in comparison we have also scaled by the drag force in an isotropic, uncharged plasma, which is represented by the black, dotted line and which we denote by $F_0=F(T,v,a=0,q=0)$. Let us first concentrate on $F_x$. We find that when $q=0$, $F_x$ starts from $F_0$ and then gradually decreases with $a$ for all the values of the velocity considered. This is in perfect agreement with the results reported earlier in \cite{Chernicoff:2012iq}. Once we turn on a non-zero value of $q$, $F_x$ starts from a much higher value at $a=0$, which is attributed to the 
presence of the charge. But then the effect of $a$ takes control, and $F_x$ diminishes steadily with increasing $a$. While the rate of decrease is intially steep, it  becomes softer at large values of $a$. $F_x$ crosses $F_0$ at some critical value of $a=a_0$, which will, in general, depend upon both $q$ and $v$. We find that the smaller the velocity, the sooner will $F_x$ come below $F_0$. Moreover, at large values of $a$ the curves corresponding to $q=0$ and $q=1$ come close together (for each value of $v$) indicating that at large values of anisotropy, the effect of charge ceases to have much significance. Coming to $F_z$ one finds that in the absence of any charge, $q=0$, $F_z$ increases monotonically with $a$ for any value of $v$. While for small values of $v$ it increases mildly, for higher values of $v$, the increase is much severe. However, the moment one introduces a finite charge density $q$ into the system it leads to an intriguing  effect. We observe that in a charged plasma $F_z$ ($a=0$) 
initially starts off with a much higher value compared to $F_0$, an effect attributed to $q$. But then for small values of $a$, $F_z$ starts to decrease until at a critical value of $a=a_m$ it attains a minima $F_{z(min)}$. $F_{z(min)}$ will depend upon the value of $v$ - for larger $v$, the minima is attained at a smaller $a_m$. However, the value of $F_{z(min)}$ is smaller for lower values of $v$. Beyond $a_m$, $F_z$ continues to rise closely following the pattern of $F_z (q=0)$ - slowly for smaller $v$ and violently for larger $v$.\newline
\noindent Let us next come to Fig.\ref{Fq} which shows the variation of $F_x$ and $F_z$ with $q$ for different values of $v$ and $a$. Here we have considered three values of $a=5,10,15$. The continuous lines stand for $v=0.9$ whereas the dotted lines stand for $v=0.7$. Again we have scaled our results by $F_0$, denoted by the black dotted line,  for convenience in comparison. Let us first concentrate on $F_x$. In the absence of anisotropy, $F_x$ starts from $F_0$ and from there on increases significantly with increasing $q$, which is shown by the black, continuous  curve (corresponding to $v=0.9$). In the presence of anisotropy, $F_x$ still increases, but now the rate of increase is toned down, with the suppression being more for greater values of $a$. Further, now one can identify a critical value of $q=q_0$ such that for $q< q_0$, one has $F_x <F_0$, on the other hand, for $q>q_0$ one has $F_x > F_0$. This is expected since we found earlier that anisotropy pushes $F_x$ down below $F_0$. It is then only 
natural that this effect will be present in the presence of charge too, until, the effect of $q$ becomes strong enough to again pull $F_x$ above $F_0$. The value of $q_0$ will depend both upon $a$ and $v$. It is found that $q_0$ increases with increasing $a$ and also with decreasing $v$. Next we move over to $F_z$ in Fig.\ref{Fq}(b). The black, continuous curve shows the variation with $q$ when $a=0$ and $v=0.9$. As one introduces anisotropy, $F_z$ increases compared to $F_0$ even at $q=0$ and for any velocity. Then it continues to rise with increase in $q$, but surprisingly, the rate of increase is now much subdued.\\
\begin{figure}[tbh] 
\begin{center}
\subfigure[]{
\includegraphics[width=7.5cm,height=6.3cm, angle=-0]{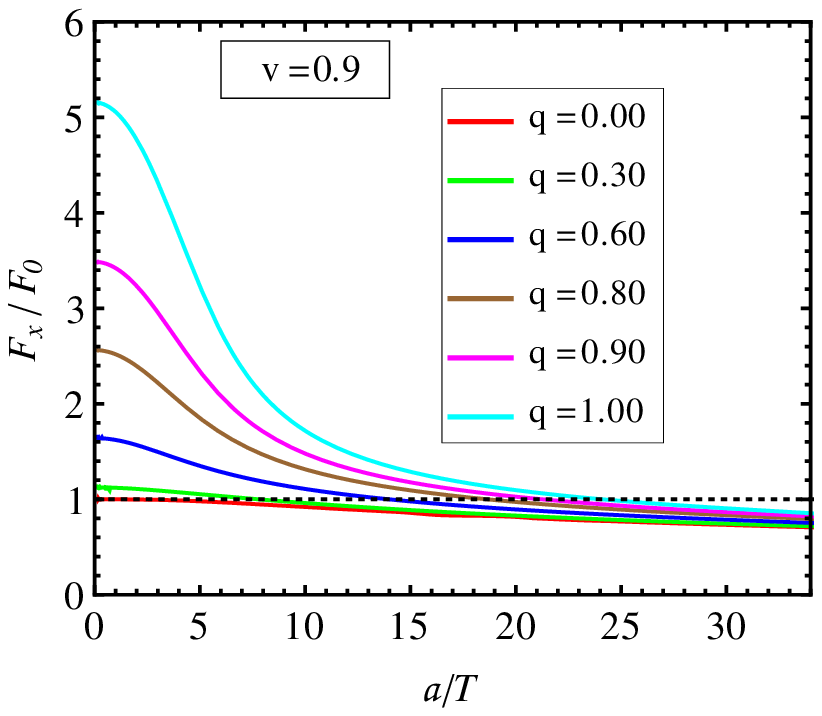}}
\hspace{5mm}
\subfigure[]{
\includegraphics[width=7.3cm,height=6.1cm, angle=-0]{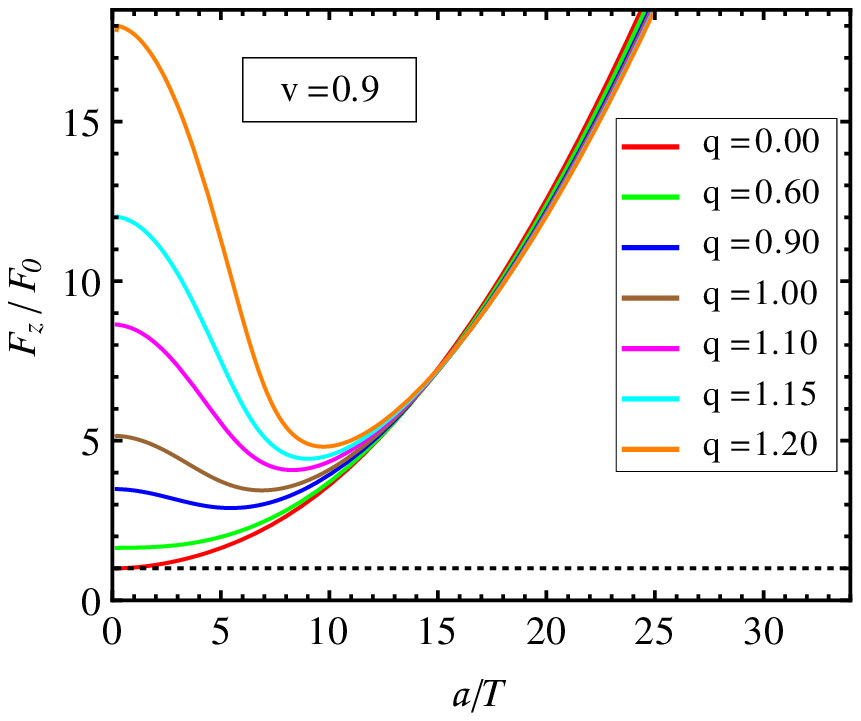}}
\caption{\label{Famanyq} \small{Variation of $F_x$ (a) and $F_z$ (b) with $a/T$ for fixed velocity $v=0.9$ and different values of $q$.}}
\end{center}
\end{figure}
In Fig.\ref{Famanyq} we again plot both $F_x$ and $F_z$ against the anisotropy parameter $a$ (scaled down by $T$), but now for a fixed $v=0.9$ and different values of the charge density $q$. As usual, the black curve stands for $F_0$. As one gradually increases $q$, $F_x(a=0)$ rises over $F_0$. But then for non-zero $a$, its effect starts to be felt. Consequently, $F_x$ decreases for all the values of $q$ considered, crosses $F_0$ at $a=a_0$ and then slowly tapers away. Again, the exact value of $a_0$ will depend upon $q$ - a smaller $q$ results in a smaller $a_0$. Also note that when $a$ is large, $F_x$ for different $q$'s almost merge indicating that the effect of $a$ reigns over that of $q$. In $F_z$ also we find that with $q$ turned on, $F_z$ begins with a much higher value compared to $F_0$. In particular, for higher $q$, $F_z(a=0)$ will also be higher. Then we find a classic instance of the interplay between $a$ and $q$. For small values of $q \sim 0.6$, $F_z$ continues to rise while for larger values, 
e.g. $q \sim 0.9$ onwards, $F_z$ decreases to reach a minima $F_{z(min)}$, beyond which it again rises. Of course, for different $q$, the minima $F_{z(min)}$ is attained at different $a_m$. Ultimately, $F_z$ for all the values of $q$ almost merge at around $a \sim 15$ and then continues to rise implying that the effect of $q$ is insignificant at large $a$. Of course, we expect the value of $a$ where all the $F_z$ curves, corresponding to different $q$'s, merge to be dependent upon the velocity $v$.\\
\begin{figure}[tbh] 
\begin{center}
\subfigure[]{
\includegraphics[width=7.4cm,height=6.1cm, angle=-0]{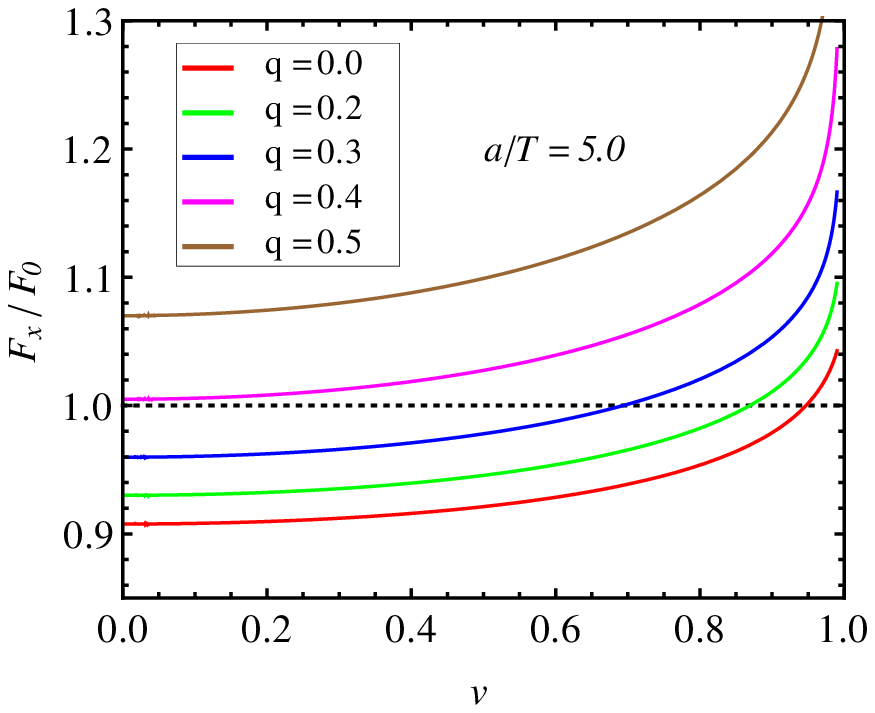}}
\hspace{5mm}
\subfigure[]{
\includegraphics[width=7.4cm,height=6.1cm, angle=-0]{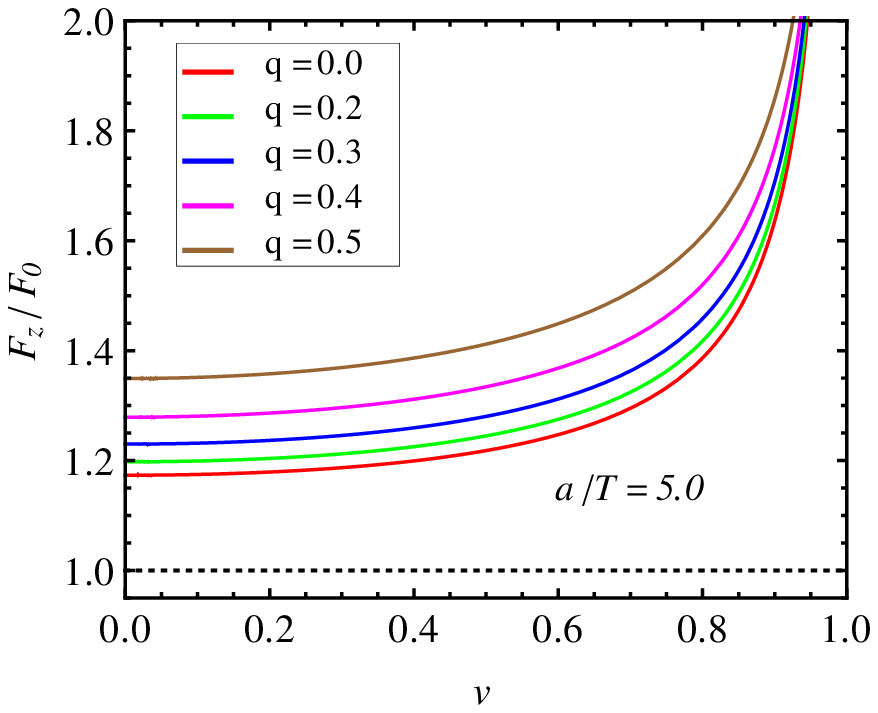}}
\caption{\label{Fv} \small{Variation of $F_x$ (a) and $F_z$ (b) with $v$ when $a/T=5$ and $q$ takes different fixed values.}}
\end{center}
\end{figure}
Fig.\ref{Fv} shows the variation of the drag force (scaled by $F_0$) with the velocity $v$ when the value of the anisotropy parameter $a/T$ and the charge density $q$ are kept fixed. We have kept the ratio  $a/T$ fixed at $5$ and considered a set of values of $q$. From Fig.\ref{Fv}(a) we find that initially when $q=0$, $F_x<F_0$. It then  increases with $v$ to cross $F_0$ at some value of $v=v_0$ determined by the specific value of $a/T$. As one puts in a small value of the charge density $q$, the initial value of $F_x$ increases, though still remaining below $F_0$ but it now crosses $F_0$ at a smaller value of $v_0$. Finally, as $q$ is raised even further, $F_x$ becomes greater than $F_0$ to start with and then continually rises. The same pattern is noticed in the case of $F_z$ except that now $F_x> F_0$ for all the  values of $q$ considered and throughout the range of $v$. Further, it is to be noticed that at large $v$, $F_z$ becomes practically independent of $q$. We shall also show this result 
analytically 
in the small anisotropy and small charge density limit to be discussed shortly.\\ 
\begin{figure}[tbh] 
\begin{center}
\subfigure[]{
\includegraphics[width=7.4cm,height=6.1cm, angle=-0]{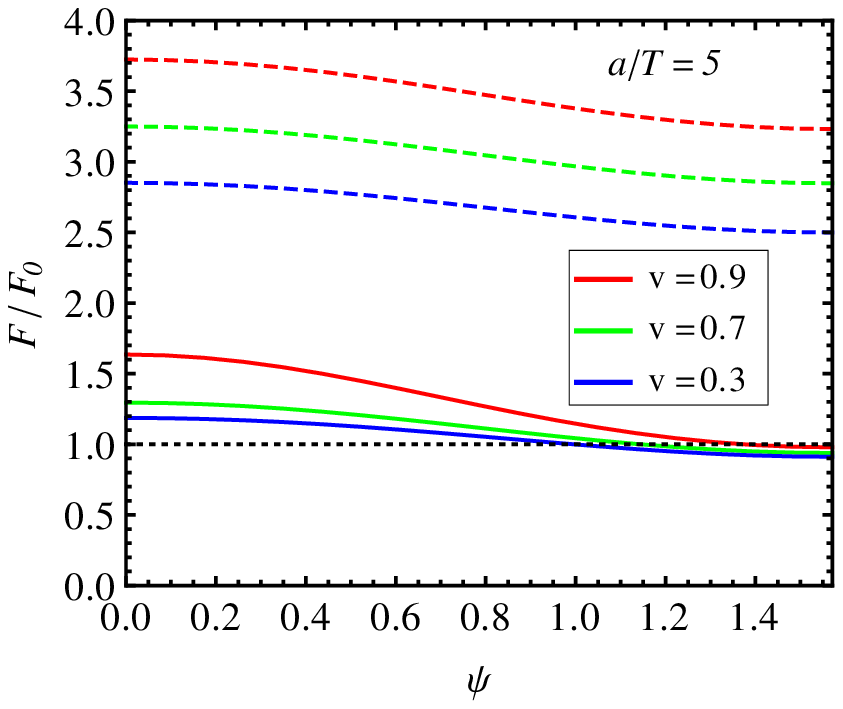}}
\hspace{5mm}
\subfigure[]{
\includegraphics[width=7.4cm,height=6.1cm, angle=-0]{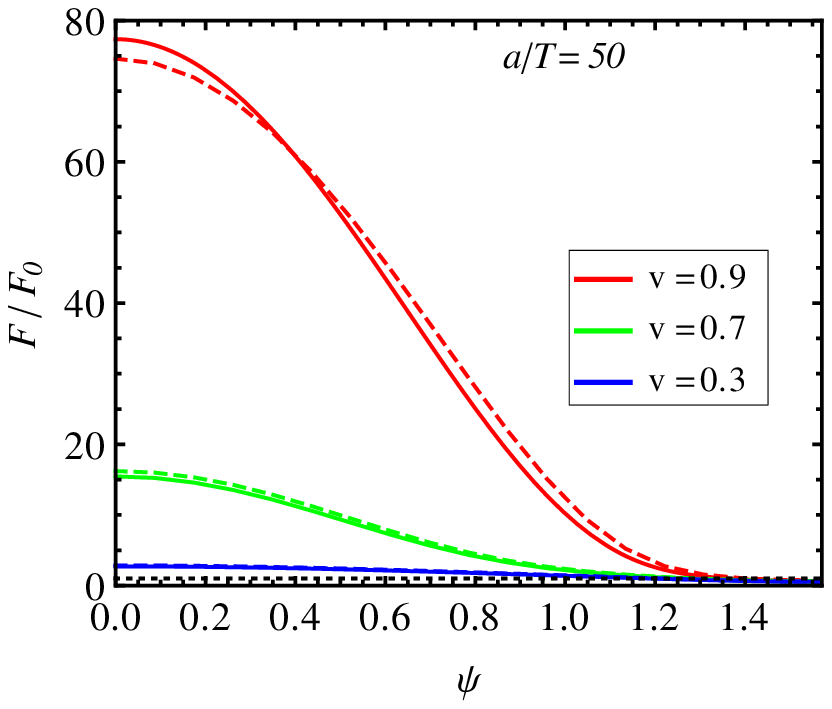}}
\caption{\label{Fpsi} \small{Variation of the drag force $F$ with the direction $\psi$ of the quark velocity (with respect to anisotropic direction) for different sets of $v$, at fixed $a$ and $q$.}}
\end{center}
\end{figure}
Fig.\ref{Fpsi} shows how the drag force $F$ varies with the direction of the quark's velocity with respect to the anisotropy axis. We have considered two values of $a/T=5,50$ respectively in Fig.\ref{Fpsi}(a) and Fig.\ref{Fpsi}(b). We have also considered $v=0.3,0.7,0.9$ and $q=0,1$ represented by continuous and dashed curves respectively. The direction of anisotropy has $\psi=0$ while $\psi=\pi/2$ represents the transverse plane. Note that, in general, the drag force along the anisotropic direction is greater than that in the transverse plane, irrespective of the values of $v,a$ and $q$. For moderate values of anisotropy, the drag force witnesses a small decrease with increasing $\psi$. But when the anisotropy parameter $a$ is large enough (as in Fig.\ref{Fpsi}(b)) the slump in the drag force is more violent. Fig.\ref{Fpsi}(a) shows that for moderate value of $a$ and $q=0$ the drag force is sensitive to the velocity for small $\psi$ whereas for large $\psi$ the drag force becomes effectively independent of 
$v$. On the other hand, for $q=1$, the velocity-dependence of the drag force persists throughout the allowed range of $\psi$ and the curves corresponding to different values of $v$ are equidistant from each other for any direction of the quark velocity. Another important observation is that as $\psi$ increases, for $q=1$ the drag force is always greater than $F_0$. But when the plasma is uncharged, the drag force becomes less than its isotropic, uncharged counterpart at a particular $\psi=\psi_0$. Thus, as the projection of the quark velocity in the transverse plane increases, at a certain $\psi_0$, that will, in general, depend upon $v$ and $a$, the drag force decreases below $F_0$. 
On the contrary, when the anisotropy is strong enough, one notes from Fig.\ref{Fpsi}(b) that while the drag force is highly sensitive to the velocity for small $\psi$, the charge density has a meager effect on the drag force. This is evident from the fact that the curves for $q=0$ and $q=1$ merge with each other. Moreover, for large $\psi$, i.e., when the quark velocity has a significant component in the transverse plane, the drag force becomes effectively independent of the velocity and the charge density. In this case, the drag force becomes smaller than $F_0$ only for large $\psi$.\\
\begin{figure}[tbh] 
\begin{center}
\subfigure[]{
\includegraphics[width=7.4cm,height=6.1cm, angle=-0]{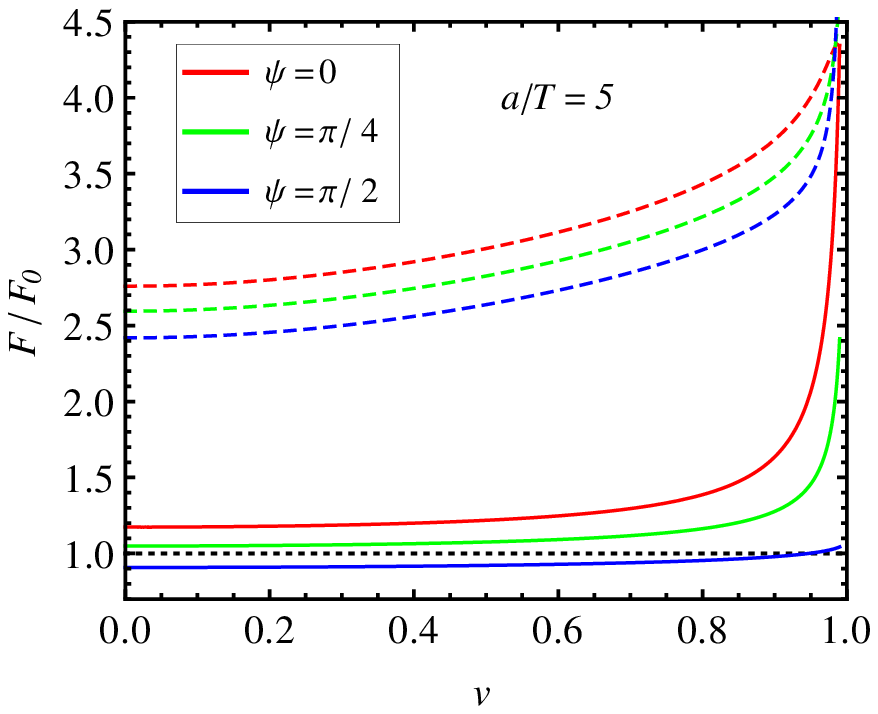}}
\hspace{5mm}
\subfigure[]{
\includegraphics[width=7.4cm,height=6.1cm, angle=-0]{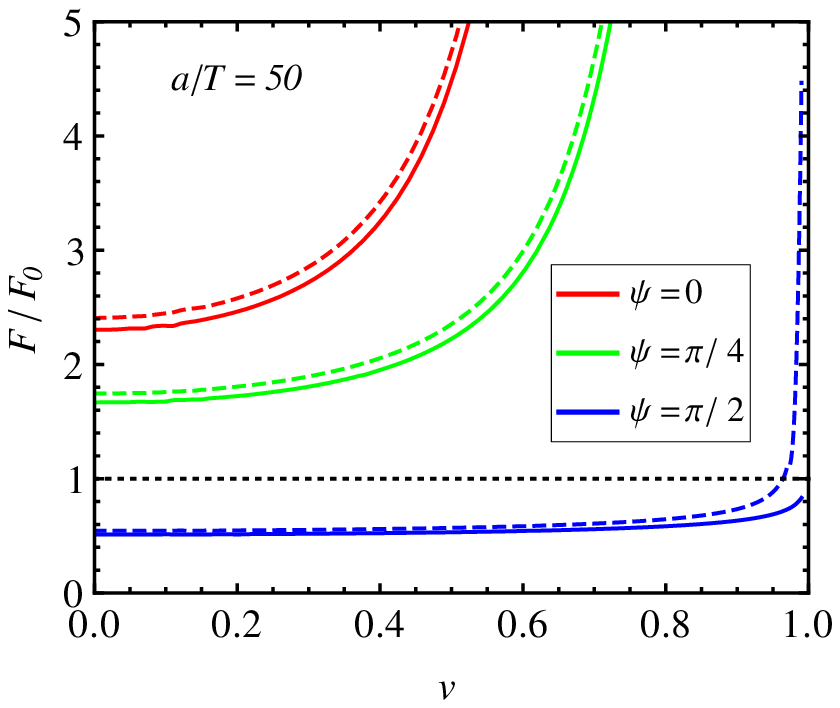}}
\caption{\label{Ftot} \small{Variation of the drag force $F$ with $v$ for different orientations of the quark velocity and $q=0,1$.}}
\end{center}
\end{figure}
Fig.\ref{Ftot} shows how the  force varies with $v$ along different directions. We have considered three cases where $\psi=0,\pi/4,\pi/2$. We have also taken $q=0,1$, indicated respectively by continuous and dashed lines,  to capture the effect of charge density. Further, we consider two values of $a/T$. First of all, observe from Fig.\ref{Ftot}(a) that in the uncharged case $q=0$ and moderate $a/T \sim 5$, the drag force increases mildly with $v$ for any direction of the quark velocity, up to a certain value of $v$, beyond which the rise in drag force becomes very steep. Also, when the motion is confined in the transverse plane, the drag force remains less than $F_0$ up to a certain value of $v$. When the plasma is endowed with a finite charge density $q$, the drag force is initially dependent upon the direction of the quark velocity, but as the ultra-relativistic regime $v  \ra 1$ is approached, $F$ becomes independent of $q$. Fig.\ref{Ftot}(b) shows that for sufficiently large $a/T \sim 50$, $F$ increases 
- initially slowly and then sharply, with $v$. While the drag force in different directions are well resolved, one notes that the curves for $q=1$ almost mimic those for $q=0$, i.e., $F$ is largely insensitive to the effect of a finite charge density.

\subsection{Small anisotropy and small charge limit analysis} \label{smallaq}
In the preceding section we have provided numerical results for the drag force for arbitrary values of the anisotropy parameter $a$ and the charge density $q$. Here we consider only small values of $a$ and $q$ and furnish some analytical results. While  restricting to small anisotropy is mandatory in order to proceed analytically, the small $q$ restriction is imposed only to make things simple. Having furnished the relevant functions earlier in the small $a$ and small $q$ regime let us proceed straight away to the results. The drag force along the $x^1$ direction, $F_x$ is given in this limit by,
\bea
F_x(T,v,a,q)\!\!\!&=\!\!\!&\f{\pi}{2}\sqrt{\la}T^2\f{v}{\sqrt{1-v^2}}\Big[1+\frac{q^2  \left(-1+v^2+3 \sqrt{1-v^2}\right)}{2 \sqrt{1-v^2}} +\frac{a^2 }{24 \pi^2 T^2\left(1-v^2\right)} \times \Big. \nn \\
&~&\Big. \left[1-v^2 +\sqrt{1-v^2}+\left(4 v^2-5\right) \log\left(1+\sqrt{1-v^2}\right)\right]\Big] \nn \\
&=&F_0 \Big[1+\frac{q^2  \left(-1+v^2+3 \sqrt{1-v^2}\right)}{2 \sqrt{1-v^2}} +\frac{a^2 }{24 \pi^2 T^2\left(1-v^2\right)} \times \Big. \nn \\
&~&\Big. \left[1-v^2 +\sqrt{1-v^2}+\left(4 v^2-5\right) \log\left(1+\sqrt{1-v^2}\right)\right]\Big]
\eea
where we have ignored terms of the $\mc{O}(a)^4, \mc{O}(q)^4$ and $\mc{O}(aq)^2$ and higher orders. In the non-relativistic limit it simplifies further as,
\bea
F_x(T,v,a,q)&=&F_0\Big[1+q^2\left(1+\f{v^2}{4}+\mc{O}(v)^4 \right) \Big. \nn \\
&~& \Big.+\f{a^2}{24 \pi^2 T^2}\left(2-5\log2+ \f{7-4\log2}{4}v^2 +\mc{O}(v)^4 \right)\Big].
\eea
In the opposite regime, when the velocity is very high, $v \sim 1$, one can write $\e(v)=1-v^2$ and the drag force assumes the form,
\be 
F_x(T,v,a,q)=F_0\left[1+\f{q^2}{2}\left(3-\sqrt{\e(v)}\right)+\f{a^2}{16 \pi^2 T^2}\left(1-\f{26}{9}\sqrt{\e(v)} \right) \right].
\ee
As alluded to earlier, the effect of $a$ is to suppress $F_x$ compared to $F_0$ whereas the effect of $q$ is to increase $F_x$ above $F_0$, which leads to an interplay when both the parameters are turned on. This enables us to identify a regime of $a$ and $q$ for which $F_x<F_0$ whereas in the complementary domain of the $a$-$q$ space $F_x > F_0$. The two domains are demarcated by a line in the $a$-$q$ plane where $F_x=F_0$. In the small $a$, small $q$ limit, this transition line obeys the equation,
\be 
q(a)=\f{1}{2\pi}\f{\sqrt{1-v^2+\sqrt{1-v^2}+(4 v^2-5)\log \left(1+\sqrt{1-v^2 }\right)}}{ \sqrt{(-1+v^2)(9-3\sqrt{1-v^2})}}\f{a}{T}
\ee
Again, in the non-relativistic limit $v \ll1$, it admits an expansion,
\be 
q(a) = \f{a}{\pi T}\left[\f{\sqrt{5\log2-2}}{2\sqrt{6}}-v^2 \f{5+\log2}{16 \sqrt{6(5\log2-2)}}+\mc{O}(v)^4\right]
\ee
which shows the transition line to be a straight line with a slope that depends upon $v$. \\
In a similar fashion one can obtain an expression for the drag force $F_z$ along the anisotropic direction in the small $a$ and small $q$ limit
\bea
F_z(T,v,a,q)\!\!\!&=\!\!\!&F_0\Big[ 1+q^2 \f{4(1+\sqrt{1-v^2})+v^2(\sqrt{1-v^2}-1)}{2 (1+\sqrt{1-v^2})^2}\Big.  \\
&~& \Big. +\f{a^2}{T^2}\f{1-v^2+\sqrt{1-v^2}+(1+v^2)\log(1+\sqrt{1-v^2})}{24 \pi^2 (1-v^2)}  \Big]+\mc{O}(a^4,q^4,a^2q^2). \nn
\eea
As in the case of $F_x$, here too we can consider two simplifying limits. In the non-relativistic limit $v \ll1$ the above expression reduces to,
\bea
F_z&=&F_0\Big[1+q^2\left(1+\f{v^2}{4}+\mc{O}(v)^4\right)\Big. \nn \\
&~&\Big.+\f{a^2}{24 \pi^2 T^2}\left(2+\log2+v^2\left(\f{1+8\log2}{4}\right)+\mc{O}(v)^4 \right) \Big].
\eea
In the ultra-relativistic limit $v \sim 1$, as before we introduce $\e(v)=1-v^2$, whence the expression for $F_z$ simplifies to,
\be 
F_z=F_0\left[ 1+\f{q^2}{2}\left(3-\sqrt{\e(v)} \right) +\f{a^2}{8 \pi^2 T^2}\left(\f{1}{\sqrt{\e(v)}}-\f{\sqrt{\e(v)}}{9} \right)\right].
\ee
Observe that as $v \ra 1$, $\e \ra 0$ and the ratio $F_z/F_0$ diverges as $(\sqrt{\e})^{-1}$. This is in agreement with the results of \cite{Chernicoff:2012iq} where the authors report that unless the quark moves in the transverse plane, the above-mentioned ratio diverges. Also observe that  at large $v$ the term depending upon $a^2$ dominates over the one depending upon $q^2$ so that in the large velocity regime, the drag force becomes practically independent of the charge density. This corroborates our earlier conclusion based on numerical analysis.

\section{Discussion and conclusion} \label{conc}
Finally, let us close with a summary of the results obtained. In this paper, we focused on the drag force on a massive quark moving through a strongly coupled hot, anisotropic plasma that carries a $U(1)$ charge density. We employed holographic techniques to explore the combined effect of anisotropy and charge density upon the drag force. While the effects of anisotropy and finite charge density were well-known previously, in this paper we explore the interplay between these parameters and uncover some interesting features when both these effects act in conjunction. While the effect of charge is to increase the drag force, the effect of anisotropy is to sharply raise the drag force along the direction of anisotropy and suppress the same in the transverse plane, in general. When both the effects act in unison, we find that the drag force in the transverse plane can become greater than its isotropic, uncharged counterpart in some regime of the parameter space.  A rather remarkable 
observation is that while increasing $a$ or $q$ cause the drag force to shoot up, in the combined presence of both these effects, the rate of increase of the drag force along the anisotropic direction is mellowed down drastically. We also find that for strong anisotropy and when the quark trajectory is confined in the transverse plane, the drag force (scaled by $F_0$) is practically independent of the velocity and the charge density. On the other hand, for large anisotropy, the drag force (scaled) along the anisotropic direction is independent of $q$ but is affected by the velocity. In the lack of much analytical control, most of our conclusions hinged upon numerical plots. However, in the limit of weak anisotropy and small charge density, the computations become more amenable to analytical treatment and we were able to furnish analytical expressions for the drag force both along the anisotropic direction and in the transverse plane. We further considered the non-relativistic and  the ultra-relativistic 
limits whence the results simplified greatly. A curious feature to come up is that in the presence of both anisotropy and charge density there exists a domain of values of $a$ and $q$ in which  the drag force in the transverse plane is smaller than the isotropic, uncharged one $F_0$, whereas in the complementary domain the drag force is greater than $F_0$. The two domains are separated by a transition line in the $a$-$q$ parameter space along which $F_x=F_0$. We were also successful in giving the functional dependence of $q$ upon $a$ along which this transition takes place. 

\appendix
\section{Details of the numerical scheme used} \label{app}
The numerical scheme we follow to find out the various functions is primarily based on the one used in \cite{Cheng:2014qia} excepting certain minor differences which we elucidate below. The relevant equations of motion are given by \cite{Cheng:2014qia}
\bea
\mc{F}&\!\!\!=\!\!\!&\frac{e^{-\frac{1}{2}\phi}}{12(\phi'+u\phi'')}\left(3a^2 e^{\frac{7}{2}\phi}(4u+u^2\phi')+48\phi'-2e^{\frac{5}{2}\phi}Q^2u^6\phi'\right) \label{eq_F_app},\\
\frac{\mc{B}'}{\mc{B}}&\!\!\!=\!\!\!& \frac{1}{24+10 u\phi'}\left(24\phi'-9u\phi'^2+20u\phi''\right),\label{eq_B_app} \\
 0&\!\!\!=\!\!\!&\frac{-48 \phi '^2 \left(32+7 u \phi '\right)+768 \phi ''+4 e^{\frac{5 \phi }{2}} Q^2 u^5 \left(-24 \phi '+u^2 \phi
 '^3-8 u \phi ''\right)}{48 \phi '-2 e^{\frac{5 \phi}{2}} Q^2 u^6 \phi '+3 a^2 e^{\frac{7 \phi}{2}} u \left(4+u \phi '\right)}
 \nonumber\\
&+&\!\!\!\frac{1}{u \left(12+5 u \phi '\right) \left(\phi '+u \phi''\right)}\Big[13u^3\phi'^4+u^2\phi'^3(96+13u^2\phi'')+8u(-60\phi''+11u^2\phi''^2-12u\phi''') \nonumber\\
 &+&\!\!\!2u\phi'^2(36+53u^2\phi''-5u^3\phi''')+\phi'(30u^4\phi''^2-64u^3\phi'''-288+32u^2\phi'')\Big].
 \label{eom}
\eea
After inflicting the transformation
\bea \label{phitrans}
\tilde{\phi}&=&\phi+\f{4}{7}\log a
\eea
as done in \cite{Mateos:2011tv, Cheng:2014qia} the differential equation for $\tilde{\phi}$ looks like
\bea
0&\!\!\!=\!\!\!&\frac{-48 \tilde{\phi} '^2 \left(32+7 u \tilde{\phi} '\right)+768 \tilde{\phi} ''+4 e^{\frac{5 \tilde{\phi} }{2}} a^{-\f{10}{7}}{Q}^2 u^5 \left(-24 \tilde{\phi} '+u^2 \tilde{\phi}
 '^3-8 u \tilde{\phi} ''\right)}{48 \tilde{\phi} '-2 e^{\frac{5 \tilde{\phi}}{2}}a^{-\f{10}{7}} {Q}^2 u^6 \tilde{\phi} '+3 e^{\frac{7 \tilde{\phi}}{2}} u \left(4+u \tilde{\phi} '\right)}
 \nonumber\\
&+&\!\!\!\frac{1}{u \left(12+5 u \tilde{\phi} '\right) \left(\tilde{\phi} '+u \tilde{\phi} ''\right)}\Big[13u^3\tilde{\phi}'^4+u^2\tilde{\phi}'^3(96+13u^2\tilde{\phi}'')+8u(-60\tilde{\phi}''+11u^2\tilde{\phi}''^2-12u{\tilde\phi}''') \nonumber\\
&+&\!\!\! u\tilde{\phi}'^2(36+53u^2\tilde{\phi}''-5u^3\tilde{\phi}''')+\tilde{\phi}'(30u^4\tilde{\phi}''^2-64u^3\tilde{\phi}'''-288+32u^2\tilde{\phi}'')\Big].
 \label{eom1}
\eea
If we further use
\bea 
\bar{Q} &=& a^{-\f{5}{7}}Q,
\eea
as done in \cite{Cheng:2014qia} we can solve the equation for $\tilde{\phi}$ in terms of $u,\tilde{\phi}_H, u_H, \bar{Q}$ using the boundary conditions at the horizon $u=u_H$ as explained in detail in \cite{Cheng:2014qia}. But to extract the effect of the charge density $Q$ (or, equivalently $q$) it is convenient to use the parameter $Q$ itself without going into $\bar{Q}$. So we solved Eq.\ref{eom1} for $\tilde{\phi}$ to get
\be 
\tilde{\phi}=\tilde{\phi}(u,u_H,\tilde{\phi}_H,Q,a).
\ee
Also from Eq.\ref{phitrans} we have 
\be \label{a}
a=e^{\f{7}{4}\tilde{\phi}(u=0,u_H,\tilde{\phi}_H,Q,a)}
\ee
where we used the fact that $\phi$ vanishes on the boundary. As R.H.S. of Eq.\ref{a} is itself a function of $a$, we can solve  it for $a$ numerically in terms of $u_H, \tilde{\phi}_H$ and which is independent of $Q$ as expected. In course of the numerical evaluation we have set $u_H=1$, which effectively means that we can use $\tilde{\phi}_H$ as an independent parameter in terms of which one can express $a$. 

\end{document}